\documentclass[fdp,a4paper,fleqn%
]{w-art}
\usepackage{times,cite
}
\usepackage[]{graphicx}
\begin{document}
\DOIsuffix{theDOIsuffix}
\Volume{55}
\Month{01}
\Year{2007}
\pagespan{1}{}



\title[Short title]{Inhomogeneities and cosmological expansion}


\author[F. Author]{Nikolaos Tetradis\inst{1,}%
  \footnote{Corresponding author\quad E-mail:~\textsf{ntetrad@phys.uoa.gr},
}}
\address[\inst{1}]{Department of Physics, University of Athens, University Campus, Zographou 157 84, Greece}
\begin{abstract}
 I review work on the influence of inhomogeneities in the matter distribution on the determination of the luminosity
distance of faraway sources, and the connection to the perceived cosmological acceleration. 
\\~\\
{\it Talk at the $9^{th}$ Hellenic School and Workshops: Standard Model and Beyond -- Standard Cosmology}
\end{abstract}
\maketitle                   





\section{An inhomogeneous model of the Universe}

In inhomogeneous cosmologies the local volume 
expansion does not necessarily coincide with the expansion rate deduced from the 
the luminosity distance of faraway sources \cite{accel}. An interesting possibility  is that the growth of
inhomogeneities in the matter distribution affects the astrophysical observations
similarly to accelerated expansion in a homogeneous Friedmann-Robertson-Walker (FRW) background. This 
may happen if the luminisoty distance is increased because of the propagation of
light through inhomogeneous regions before reaching the observer. 

At length scales above $\sim 50$ Mpc the density contrast in the Universe is at most of ${\cal O}(1)$. 
A popular modelling of the cosmological background is based on the Lemaitre-Tolman-Bondi (LTB) 
metric. This geometry has spherical symmetry, but can 
be inhomogeneous along the radial direction. Several spherical regions, described by the LTB metric, can be
embedded in a homogeneous FRW background. This construction is characterized as a LTB Swiss-cheese model.
There are two possible choices for the location of an observer, which are consistent
with the isotropy of the Cosmic Microwave Background: 
i) in the interior of a spherical inhomogeneity, near its center;
ii) in the homogeneous region, with the light travelling
across several inhomogeneities during its propagation from source to observer.

The LTB metric can be written in the form
\begin{equation}
ds^{2}=-dt^2+\frac{R'^2(t,r)}{1+f(r)}\,dr^2+R^2(t,r)d\Omega^2,
\label{metrictb}
\end{equation}
where $d\Omega^2$ is the metric of a two-sphere  and
$f(r)$ is an arbitrary function.
The function $R(t,r)$ describes the location of a shell of matter marked by $r$
at the time $t$.
The Einstein equations give
\begin{equation}
\dot{R}^2(t,r)=\frac{1}{8\pi M^2}\frac{{\cal M} (r)}{R}+f(r)
\label{tb2} \end{equation}
where ${\cal M}'(r)=4\pi R^2 \rho(t,r) \, R'$ and
$G=\left( 16 \pi M^2 \right)^{-1}$.
The generalized mass function ${\cal M}(r)$ of the pressureless fluid 
with energy density $\rho(t,r)$ can be chosen arbitrarily. 

We parametrize the energy density at some arbitrary initial time 
as $\rho_i(r)=\rho(0,r)=\left( 1+\epsilon(r)\right)\rho_{0,i}$.
The initial energy density of the homogeneous background surrounding the spherical 
inhomogeneity is
$\rho_{0,i}$. If the size of the inhomogeneity is
$r_0$, the matching with the homogeneous metric in the exterior requires
$4\pi\int_0^{r_0} r^2 \epsilon(r) dr=0$,
so that
${\cal M}(r_0)=4\pi r^3_0 \rho_{0,i}/3.$
As we assume that the homogeneous metric is flat,
we also have $f(r_0)=0$. 
The typical evolution of such an inhomogeneous background is depicted in fig. \ref{profile}. The configuration
models a void, with a central underdensity surrounded by an overdensity. The central density is reduced during
the cosmological evolution, while the matter is concentrated in the periphery. 

\begin{vchfigure}[t]
  \includegraphics[width=.7\textwidth,height=72mm]{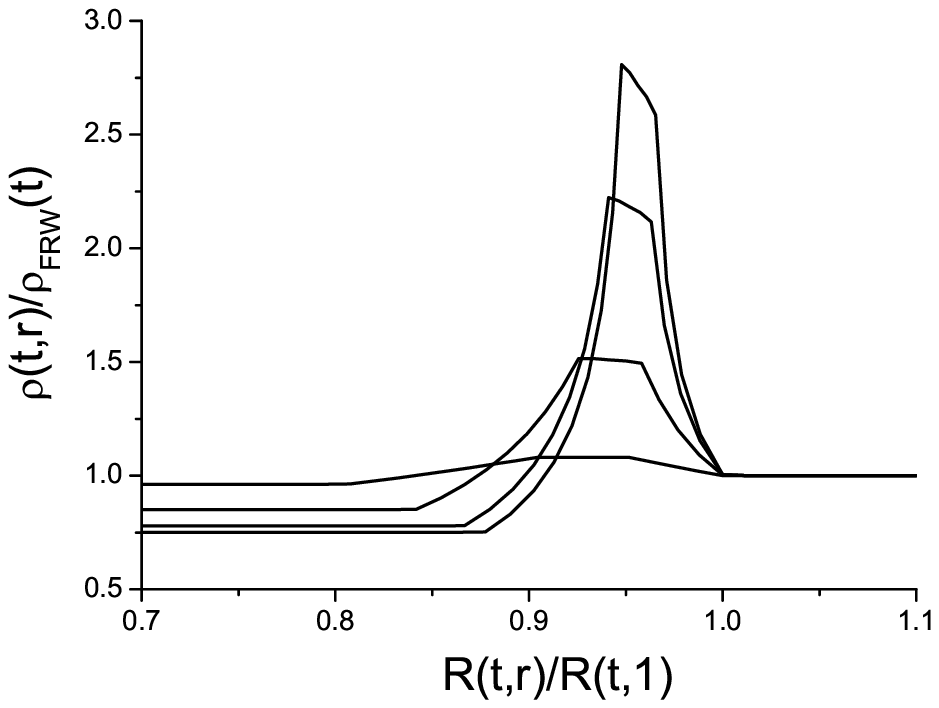}
\vchcaption{The evolution of the density profile for a central underdensity surrounded by
an overdensity.}
\label{profile}
\end{vchfigure}

\section{Propagation of light beams and luminosity distance}

The optical equations \cite{sachs} describe the evolution of the characteristics of a beam 
(area and shape of its cross-section) during its propagation in a given gravitational background.
For a LTB Swiss-cheese model, with a density contrast not much larger than 1, the relevant equation is 
\cite{brouzakis}
\begin{equation}
\frac{1}{\sqrt{A}}\frac{d^2\sqrt{A}}{d\lambda^2} =
-\frac{1}{4M^2}\rho \left( k^0\right)^2 ,
\label{exx3} \end{equation}
where $A$ is the cross section of a light beam, $\lambda$ an affine parameter
along the null trajectory and $k^i={dx^i}/{d\lambda}$.
We neglect the shear tensor, which describes deformations of the beam, 
because it is important only when the beam passes near regions in which the
density exceeds the average one by several orders of magnitude.
We assume that the light emission near the source is not
affected by the large-scale geometry. By choosing an affine parameter
that is locally $\lambda=t$ in the vicinity of the source, we can set
$\left.{d\sqrt{A}}/{d\lambda} \right|_{\lambda=0}=\sqrt{\Omega_s}$,
where $\Omega_s$ is the solid angle spanned by the beam when the light is emitted 
by a point-like isotropic source.
This relation and
$\left. \sqrt{A} \right|_{\lambda=0}=0$
provide the initial conditions for eq. (\ref{exx3}).

\begin{vchfigure}[t]
  \includegraphics[width=.7\textwidth,height=72mm]{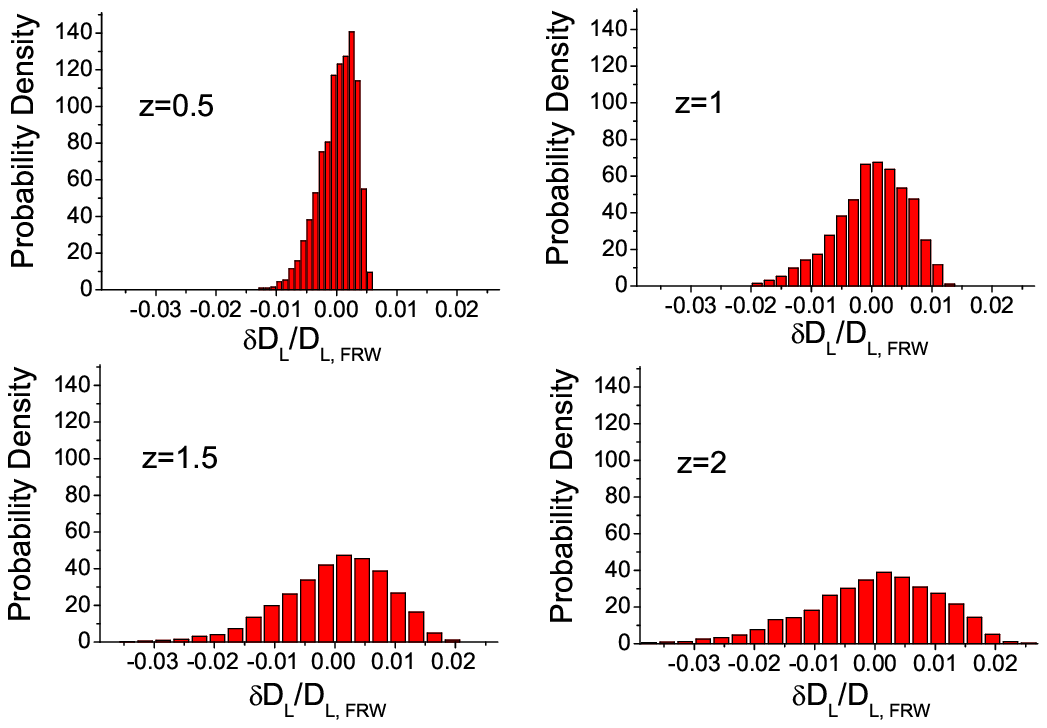}
\vchcaption{The distribution of luminosity distances for various redshifts in the LTB Swiss-cheese
model if the inhomogeneities have a characteristic scale of $133\,h^{-1}$ Mpc.}
\label{lum133}
\end{vchfigure}

In order to define the luminosity distance, we consider photons
emitted within a solid angle $\Omega_s$
by an isotropic source with luminosity $L$.
These photons are detected
by an observer for whom the light beam
has a cross-section $A_o$.
The redshift factor is
$1+z={\omega_s}/{\omega_o}={k^0_s}/{k^0_o}$.
The luminosity distance is 
$D_L=(1+z)\sqrt{A_o/\Omega_s}$, with $A_o$ the beam area measured 
by the observer
for a beam emitted within $\Omega_s$. 
The beam area can be calculated by solving
eq. (\ref{exx3}).
We consider light beams that pass through several inhomogeneities.
The light is emitted from a point
at the edge of the first inhomogeneous region, with a random initial direction, and
moves through it.
Subsequently, the beam crosses the following inhomogeneity in a similar
fashion. The angle of entry into the new inhomogeneity is assumed again to be
random. The initial conditions are set by the values of
$\sqrt{A}$ and $d\sqrt{A}/{d\lambda}$ at the end of the first crossing. This process
is repeated until the light arrives at the observer.

In fig. \ref{lum133} we depict the distributions of
the deviation of the luminosity distance from the value in
a homogeneous background
for various redshifts \cite{brouzakis,brouzakis2}. The figure corresponds to a Swiss-cheese model with 
inhomogeneities of a common length scale of $133\,h^{-1}$ Mpc.
The total integral of
the distributions has been normalized to 1 in all cases, so that they are in fact
probability densities.
They have similar profiles that are asymmetric around
zero. Each distribution has a maximum at a value larger
than zero and a long tail towards negative values. The average deviation is zero to
a good approximation in all cases. This is expected because of flux conservation \cite{weinberg,brouzakis2}:
As long as the light propagation in
an inhomogeneous background does not modify significantly the redshift, the
energy may be redistributed in various directions through gravitational lensing by inhomogeneities, 
but the total flux is conserved and
remains the same as in a FRW background.

The longer tail of the distribution towards small luminosity distances 
is a consequence of the presence of a thin and dense spherical
shell around each central underdensity. The portion of light beams that
cross several shells is small. However, the focusing is
substantial for such beams and the resulting luminosity distance much shorter than
the average. The effect of the long tail is compensated by the shift of the
maximum of the distribution towards positive values.
The form of the distribution is very similar to that derived in studies modelling
the inhomogeneities through the standard Swiss-cheese model \cite{holz}.
In that case the strong focusing is generated by the very dense concentration of matter
at the center of each spherical inhomogeneity. We emphasize, however, that the two models
have a different region of applicability. The standard Swiss-cheese model \cite{holz} is appropriate
for length scales of ${\cal O}(1)\, h^{-1}$ Mpc or smaller, while the LTB Swiss-cheese model \cite{brouzakis,brouzakis2}
for scales of ${\cal O}(10)\, h^{-1}$ Mpc or larger.

The width of the distribution $\delta_d$
determines the error induced to cosmological parameters
derived through the luminosity curve,
while the location of its maximum $\delta_m$ the bias in such determinations. 
A small sample of data is expected to favour
values of the luminosity distance near the maximum of the distribution, and thus
generate a bias \cite{holz,brouzakis2}. An important quantity is the effective equation 
of state $w=p/\rho$ deduced from astrophysical data. 
The presence of inhomogeneities induces a
statistical error in $w$ , as well as a shift of its
average value if the sample is small \cite{brouzakis2}. 
For inhomogeneities with a typical size of  $40\,h^{-1}$ Mpc the error
is $\delta w\simeq 0.015$ for all $z$ between 0.5 and 2, while the average value $\bar{w}$
for a small sample is negative and of ${\cal O}(10^{-3})$.
For a size of $133\,h^{-1}$ Mpc
the error increases from 0.015 to 0.025 as $z$ increases from
0.5 to 2, while the average value $\bar{w}$ is again negative and of ${\cal O}(10^{-3})$. 
For a size of $400\,h^{-1}$ Mpc
the error increases from 0.03 to 0.05 for $z$ increasing from 0.5 and 2, while the average
is $\bar{w} \simeq -0.015$. 
The values of $\delta_d$ and $\delta_m$ can be compared to those generated by the
effects of gravitational
lensing at scales characteristic of galaxies or clusters of galaxies (modelled through the standard
Swiss-cheese model) \cite{holz}.
The typical values of $\delta_d$ and $\delta_m$ are larger by at least an
order of magnitude than the ones we obtained.
The reason is the difference in the density contrast.

We conclude that, if the source and the observer have random locations, 
the presence of inhomogeneities with large length scales -
even comparable to the horizon
distance - and density contrast of ${\cal O}(1)$ does not influence 
the propagation of light sufficiently in order to explain the supernova data without dark energy.
For this to be possible the effect on $w$ would need to be close to 1.
However, the errors induced in the measurements of the luminosity distance of high-redshift sources can be substantial,
depending on the modelling of the inhomogeneous background.
Care must be taken in the extraction of cosmological parameters from such data. 

\section{Analytical estimates and a central observer}

For a smooth density field with a contrast of ${\cal O}(1)$, 
the size of an inhomogeneity $r_0$ determines its effect 
on quantities such as redshift and luminosity distance of a source. An analytical estimate of
the effect is possible \cite{estimate}.
The relevant quantity is the dimensionless ratio $\bar{H}=r_0 H$ of $r_0$ to
the horizon distance $1/H$. 

If the observer is located at a random position within the
homogeneous region, each crossing of an inhomogeneity produces an 
effect of ${\cal O}(\bar{H}^3)$ for the travel time and the redshift.
For the beam area and the luminosity distance the effect 
is of ${\cal O}(\bar{H}^2)$.  
However, flux conservation implies
that positive and negative contributions to the beam area 
cancel during multiple crossings. 
The size of the {\it maximal average} effect of each crossing on
the beam area and luminosity distance is set by the effect on
the redshift, which is of ${\cal O}(\bar{H}^3)$ \cite{weinberg,brouzakis2}. 
Photons with redshift $\sim 1$ pass through $\sim (1/H)/r_0=\bar{H}^{-1}$ inhomogeneities
before arrival, assuming that these are tightly packed. 
As a result, the expectation is that the maximal final effect for a random
position of the observer is of 
${\cal O}(\bar{H}^2)$ for all quantities. 
Allowing for corrections arising from numerical factors, this conclusion is
supported by the detailed analysis of \cite{brouzakis,brouzakis2}.  

For an observer located at the center of a spherical inhomogeneity, the
deviations of travelling time, redshift, 
beam area and luminosity distance from their values
in a homogeneous background are of ${\cal O}(\bar{H}^2)$. The luminosity 
distance is increased by the presence of a central underdensity, while it is
reduced by a central overdensity. 
The increase in the luminosity distance if the observer is 
located near the center of a large void
can by employed for the explanation of the supernova data.
An increase of ${\cal O}(10\%)$, as required by the data, would imply the 
existence of a void with size close to $10^3\, h^{-1}$ Mpc. 
Numerical factors can reduce the required size, depending on the details of
the particular cosmological model employed. However, a typical void 
with size of ${\cal O}(10)\, h^{-1}$ Mpc leads to a negligible 
increase of the luminosity distance.

\begin{acknowledgement}
 N.~T. is supported in part by the EU Marie Curie Network ``UniverseNet'' 
(HPRN--CT--2006--035863) and the ITN network
``UNILHC'' (PITN-GA-2009-237920).
\end{acknowledgement}

\end{document}